# HOME AUTOMATION

Z. Ahmed

*Abstract*— **In this paper I briefly discuss the importance of home automation system. Going in to the details I briefly present a real time designed and implemented software and hardware oriented house automation research project, capable of automating house's electricity and providing a security system to detect the presence of unexpected behavior.**

*Index Terms*— Automation, Bit Controllers, Security

## I. INTRODUCTION

TODAY the technological world's main focus is to automate every possible thing to take advantage in providing ease in life. In the most of the well developed countries home automation is very famous and well adopted because it has several many benefits like it saves electricity and provides security. In this short paper I am also briefly going to discuss automation but only related to the houses.

## II. AUTO ELECTRIC HOUSE

The Auto Electric House is a house whose electricity is controlled and security mechanism by computer. The whole project is consists of following different software and hardware components .i.e., *House with electricity, Security System, Bit controller card, Adapter (12V DC), Programmed Software.*

### A. Electric House

The prototype house has been constructed with wood and glass. The whole house is consists of six rooms and one lobby. The electricity has been provided to all of the rooms and the lobby of the house. Only one computer is controlling the whole electricity of the house. The house consists of three layers .i.e., *External Layer, Internal Layer and Deep Internal Layer*.

*1) External Layer*
This is the outer shown part of the house covering the house from upside like roof of the house. This part of the house has made of wood and glass. Only the security system has been planted at this layer. (See Figure 1)

*2) Internal Layer*
This is the middle layer of the house because this layer is placed between the other two layers of the house (Deep internal and external layer). Basically this layer carrying the internal structure of the house (rooms, lobby and some other parts). The electric components, which are automatically controlled are fixed in this part of the house. (See Figure 2)

*3) Deep Internal Layer*
This is the most important layer of the house because this contains the main hardware designed and made to control the electricity of house and provide complete security system. (See Figure 3)

### B. Security System

This is an important feature of the auto electric house. The security has been implemented on the external layer of the house and its function is to alert owner if someone disturbs the security. The security system is infrared based, when ever some disturbance occurs then it reports input to the controlling computer to perform needed actions.

### C. Bit controller card

Bit controller card operates at 12V (DC) provided by the adapter. Controller card is connected to the computer using parallel port cable. The main circuit of bit controller card is based upon 8 sub circuits and following components have been used in the circuit.

1. Relays
2. Transistors
3. Crystals Diodes
4. Resistances
5. Capacitor
6. Bridge
7. Connector
8. Parallel port

*1) Relays*
The main function of relay is to switch between two terminals. The connection of each room is connected to each relay. Operating voltage of the relay is 12 volts (DC),

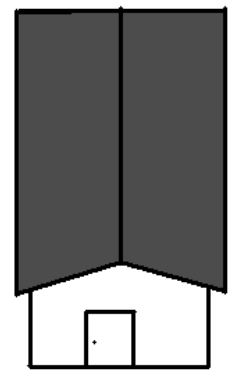

Figure .1. External Layer

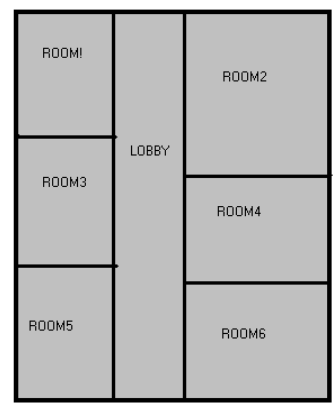

Figure .2. Internal Layer

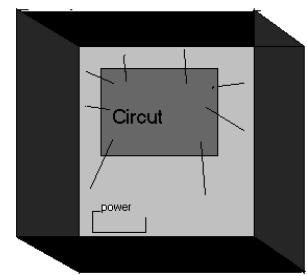

Figure .3. Deep Internal Layer

□Z. Ahmed is presently working as University Assistant with Mechanical Engineering Informatics and Virtual Product Development Division, Vienna University of Technology, Getreidemarkt 9/307 1060 Vienna Austria (phone: 004315880130726 - email: zeeshan.ahmed@tuwien.ac.at  and zeeshan.ahmed@hotmail.de).

Maximum switching current that can be passed through relay can be of 220 volts 5A (AC). Magnetic field is produced when current is passed through the coil of the relay.

During forward biasing relay connects with terminal 1 and in case of reverse biasing relay connects to the terminal 2. This reverse and forward biasing is handled with the help of a transistor. (See Figure 4 and 5)

*2) Transistors – C945*

Transistor C945 is used in this circuit to control biasing. Maximum operating voltage of a transistor is (1-1.5V, 0.3A). Moreover used resistances are connected across the transistor.

*3) Crystals diodes*

Crystal diode is used to purify the biased current flowing

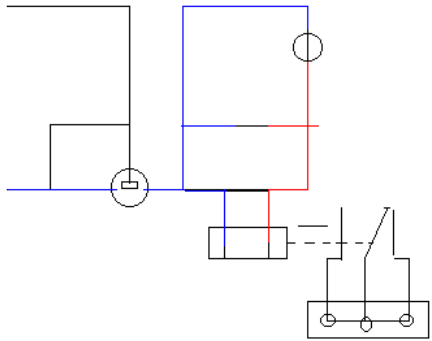

Figure .4. Forward Biasing

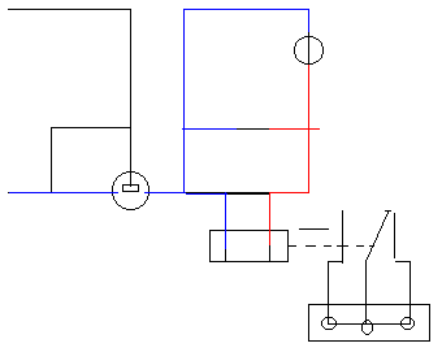

Figure .5. Reverse Biasing

toward relay.

*4) Resistances*

The main purpose of resistance is to drop flow of current up to extent of resistance. Two types of resistances are used in this circuit .i.e., 1k and 4.7k.

1k resistance is connected across LED and combination of 4.7k and 1k resistance is connected across transistor.

*5) Capacitor*

The capacitor is used to store charge and maintain constant flow of current. 25V 2200 MICRO FARAD Capacitor is used.

*6) Bridge*

A bridge is a combination of four diodes and used to convert AC voltage into DC voltage.

*7) Connector*

Connector is used to connect AC transformer of 12V,1A to the circuit.

*8) Parallel port*

A parallel port is used to connect hardware with computer. We divided parallel port in three parts (Port A, Port B, Port C).

1. Port A - consists of selection lines D0 – D7 (from connectors hole # 2 to #9).
2. Port B – consists of ERROR, SLCT, PE, ACK, BUSY lines (from hole # 15, 13, 12, 10, 11 respectively).
3. Port C – consists of AUTOFND#, INIT#, SLCTIN# lines (from hole # 14, 16, 17 respectively).

*D. Adapter*

Adapter of 220V AC is used to supply current to the circuit.

*E. Programmed Software*

The software is programmed in C language. The basic function of the software is to exchange and transmit bit based information between hardware and computer.

III. AUTO ELECTRIC HOUSE – WORK FLOW

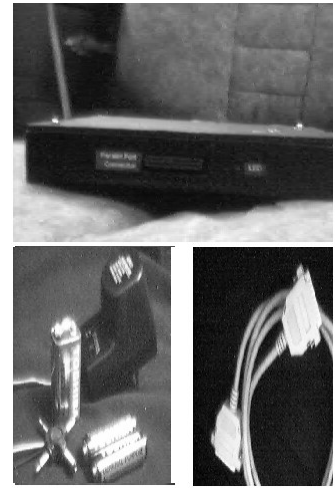

Figure .6. Hardware

The flow of current to relays is controlled by the designed circuit. The circuit is designed using the combination of transistors (C945), crystal diode and three resistances respectively (1k, 1k and 4.7k).

Whereas LED is used to indicate the status of relay whether it is switched or not.

A transformer connected to 220V (AC) produces 12V (AC), this 12V (AC) is input in the circuit through connector. Then this AC voltage across through bridge performs rectification in contact with capacitor and in result pure balanced (12V DC) current is obtained at the terminals of capacitor (−ve at −ve terminal and +ve at +ve terminal).

This current passes through transistor, forward biasing is produced, biased current is transferred to relay passing through crystal diode. Hence relay operates positively (connected to terminal (1)). (See Figure 7)

When pulse is generated from the parallel port to its respective circuit the passing current through transistor is reverse biased hence reverse current is passed to the relay so that it can connect terminal (2) and LED turns on.

## IV. Conclusion

In this short paper I have briefly discussed the importance of home automation system then provided the brief information about a PCB circuit construction to control house's electricity and provide infra red based security system connected and controlled by a computer via parallel port.

## Acknowledgement

As the part of acknowledgement I would like to thanks Prolife All Pakistan Project Exhibition for giving me the opportunity to participate and present this project in the nationwide software project exhibition and competition Pakistan.